\documentclass[12pt,a4paper]{article}
\usepackage{a4wide}
\usepackage{amssymb}
\usepackage{amsmath}
\usepackage{cite}

\usepackage[hyperindex=true,
          pdfstartview=FitH,
          bookmarksnumbered=true,
          bookmarksopen=true,
          citecolor=blue,
          linkcolor=blue,
          colorlinks=true,
          unicode]{hyperref}

\allowdisplaybreaks
\numberwithin{equation}{section}

\begin{document}

\title{The Entropy Sum of (A)dS Black Holes in Four and Higher Dimensions}
\author{
         Wei Xu$^{1,2}$\thanks{{\em
        email}: \href{mailto:xuweifuture@mail.nankai.edu.cn}{xuweifuture@mail.nankai.edu.cn}}\ ,
        Jia Wang$^{1}$\thanks{{\em
        email}: \href{mailto:wangjia2010@mail.nankai.edu.cn}{wangjia2010@mail.nankai.edu.cn}}\, and
        Xin-he Meng$^{1,3}$\thanks{{\em
        email}: \href{mailto:xhm@nankai.edu.cn}{xhm@nankai.edu.cn} corresponding author}\\
$^{1}$School of Physics, Nankai University, Tianjin 300071, China\\
$^{2}$School of Physics, Huazhong University of Science and Technology, \\ Wuhan 430074, China\\
$^{3}$State Key Laboratory of ITP, ITP-CAS, Beijing 100190, China
}
\date{}
\maketitle

\begin{abstract}
We present the ``entropy sum'' relation of (A)dS charged black holes in higher dimensional Einstein-Maxwell gravity, $f(R)$ gravity, Gauss-Bonnet gravity and gauged supergravity. For their ``entropy sum'' with the necessary effect of the un-physical ``virtual'' horizon included, we conclude the general results that the cosmological constant dependence and Gauss-Bonnet coupling constant dependence do hold in both the four and six dimensions, while the ``entropy sum'' is always vanishing in odd dimensions. Furthermore, the ``entropy sum'' of all horizons is related to the geometry of the horizons in four and six dimensions. In these explicitly four cases, one also finds that the conserved charges $M$ (the mass), $Q$ (the charge from Maxwell field or supergravity) and the parameter $a$ (the angular momentum) play no role in the ``entropy sum'' relations.
\end{abstract}

\section{Introduction}
There has recently been some considerable ongoing interest in the ``universal property'' of horizon entropy for various types of multi-horizons black holes. It seems clear that these additional thermodynamic relations of entropy may provide further insight into the quantum physics for black holes. One recent interest has been focused on the ``area product'' or ``entropy product'' rules
\cite{Cvetic:2010mn,Castro:2012av,Visser:2012wu,Chen:2012mh,Toldo:2012ec,
Faraoni:2012je,Detournay:2012ug,Visser:2012zi,Cvetic:2013eda,Abdolrahimi:2013cza,Lu:2013ura,Anacleto:2013esa,
Chow:2013tia,Castro:2013kea,Castro:2013pqa,Lu:2013eoa}, which is often independent of the mass of black holes
\cite{Cvetic:2010mn,Castro:2012av,Toldo:2012ec,Chen:2012mh,Visser:2012zi,Cvetic:2013eda,Abdolrahimi:2013cza,Lu:2013ura,
Anacleto:2013esa,Chow:2013tia,Castro:2013kea,Lu:2013eoa} and depends only on various charges and angular momentums mostly. But the mass independence claim sometimes fails \cite{Faraoni:2012je,Castro:2013pqa,Detournay:2012ug,Visser:2012wu}.
Another ``universal property'' of entropy is the ``entropy sum'' relation of black holes \cite{Wang:2013smb,Xu:2014qaa,Wang:2013nvz}. It is shown that, in the four dimensional Einstein-Maxwell-(A)dS spacetime, the ``entropy sum'' depends only on the cosmological constant with  the necessary effect of the un-physical ``virtual'' horizon included. When in the
spacetime with extra scalar field and in Einstein-Weyl spacetime, they find the ``entropy sum''  is dependent of the constants characterizing the strength of the extra degree of freedom, the matter field. Besides, the ``entropy sum'' relation does not depend on the conserved charges $M$ (mass), $Q$ (charge from Maxwell field) and $J$ (from ``rotation field'' ). Thus, it is said that the ``entropy sum'' is also universal and is only related to the background field properties. Furthermore, it is found that entropy product and entropy sum of the
Schwarzschild-(A)dS black hole are somehow ``equal'' \cite{Xu:2014qaa}, when only the effects of the physical horizons are considered, as they both can be simplified into a mass independent entropy relation of physical horizon \cite{Visser:2012wu,Xu:2014qaa}. That is to say, the entropy sum also may somehow reveal the microscopic physics of black holes.

On the other hand, the entropy product of two-horizons black holes has been understood well and physically via their holographic description, i.e. the thermodynamic method of black hole/Conformal Field Theory (BH/CFT) correspondence \cite{Chen:2012mh,Chen:2012yd,Chen:2012ps,Chen:2012pt,Chen:2013rb,Chen:2013aza,Chen:2013qza}. By using the BH/CFT correspondence, it is found that the same central charge $c_R = c_L$ is equivalent to the condition that the entropy product $S_+S_-$ being mass-independent, or equivalently the condition $T_+S_+=T_-S_-$, where $T_{\pm}$, $S_{\pm}$ are the outer and inner horizon temperatures and entropies respectively. Therefore the entropy product $S_+S_-$ being mass-independent (equivalently thermodynamics relations $T_+S_+=T_-S_-$) may be taken as the criterion whether there is a 2 dimensional CFT dual for the black holes in the Einstein gravity and other diffeomorphism invariant gravity theories \cite{Chen:2012mh,Chen:2012yd,Chen:2012ps,Chen:2012pt,Chen:2013rb,Chen:2013aza,Chen:2013qza}. In this sense, the equality between entropy product and entropy sum leads one to expect that the entropy sum may be understood well in the similar way. This makes it more interesting for studying the entropy sum in four and higher dimension (A)dS spacetime, especially for the cases that the mass-independence of entropy product fails.

The surprising discovery of the cosmic late stage accelerating expansion has inspired intensive research on the universe background cosmological constant, including its effects on the astrophysics and black hole physics. In this paper, we generalize the discussion about the ``entropy sum'' to higher dimensions cases. We will mainly test the ``entropy sum'' relation of (A)dS charged black holes in higher dimensional Einstein-Maxwell gravity, $f(R)$ gravity and Gauss-Bonnet gravity. For their ``entropy sum'' with including the necessary effect of the
un-physical ``virtual'' horizon, we conclude the general results that the cosmological constant dependence and Gauss-Bonnet coupling constant dependence do still hold in both the four and six dimensions, while the ``entropy sum'' is always vanishing in odd dimensions. Furthermore, the ``entropy sum'' of all horizons is related to the topology of the horizons in four and six dimensions. In these clearly three cases, one also finds that the conserved charges $M$ and $Q$ play no role in the ``entropy sum'' relation.

This paper is organized as follows. In the next Section, we will investigate the ``entropy sum'' of higher dimensional (A)dS charged black holes. In Sections 3, 4 and 5 we discuss the ``entropy sum'' of (A)dS black hole in higher dimensional $f(R)$ gravity and Gauss-Bonnet gravity, gauged supergravity respectively. Section 6 is devoted to the conclusions.

\section{The ``entropy sum'' of (A)dS black hole in Einstein-Maxwell gravity}
The Einstein-Maxwell action in higher dimensions $d$ reads as
\begin{align}
  \mathcal{L}=\frac{1}{16\pi G}\int d^d x \sqrt{-g}[R-2\Lambda-F_{\mu\nu}F^{\mu\nu}],
\end{align}
where $\Lambda=\pm\frac{(d-1)(d-2)}{2\ell^2}$ is the cosmological constant associated with cosmological scale $\ell$. Here, the negative cosmological constant corresponds to AdS spacetime while positive one corresponds to dS spacetime. Varying this action with respect to the metric tensor leads to the RN-AdS solution given by \cite{Liu:2003px,Astefanesei:2003gw,Brihaye:2008br,Belhaj:2012bg}
\begin{align} \label{metric}
  d s^2=-V(r) d t^2+\frac{d r^2}{V(r)}+r^2 d\Omega^2_{d-2},
\end{align}
where $d\Omega^2_{d-2}$ represents the line element of a
$(d-2)$-dimensional maximal symmetric Einstein space with constant
curvature $(d-2)(d-3)k$, and the $k=1, 0$ and $-1$, corresponding
to the spherical, Ricci flat and hyperbolic topology of the black
hole horizon, respectively. The metric function $V(r)$ is given by
\begin{align}
  V(r)=k-\frac{2M}{r^{d-3}}+\frac{Q^2}{r^{2(d-3)}}-\frac{2\Lambda}{(d-1)(d-2)}r^2,
\label{V(r)1}
\end{align}
where $M$ and $Q$ are the mass and the charge of the black hole, respectively. In the Einstein-Maxwell gravity, the entropy of horizon located in $r=r_i$, as usual, is given by
\begin{align} \label{entropy1}
  S_i=\frac{A_i}{4}=\frac{\pi^{(d-1)/2}}{2\Gamma\left(\frac{d-1}{2}\right)}r_i^{d-2},
\end{align}
where $A_i$ is the volume of the $(d-2)$-sphere with the ``radius'' $r_i$
\begin{align*}
A_i=\frac{2\pi^{(d-1)/2}}{\Gamma\left(\frac{d-1}{2}\right)}r_i^{d-2}.
\end{align*}

Back to the horizon function Eq.(\ref{V(r)1}), in principle this high order polynomial can have at most $2(d-2)$ roots (even number), while some of them may be equal at special parameter conditions between $M, Q$ and $\Lambda$. In what follows, we will only consider the case for the $2(d-2)$ roots, as we are interested in the ``entropy sum'' of multi-horizons black hole. That is to say, we will consider all possible roots of the horizon structure in the following paper. One needs noting that, in fact we are considering the black holes with some special black hole mass $M$ in the following discussions, in order to have multi-roots or more horizons.

However, the horizon function can not always be solved explicitly. Thus, we will only take some cases with specific dimensions $d$ as examples. And one will find that at the end of day it is not at all necessary to list all the roots explicitly out. What we emphasize is that there are four kind of roots, each of which stands for the event horizon, Cauchy horizon, cosmological horizon and the un-physical ``virtual'' horizon respectively. We find the ``virtual'' horizon can not be dropped, otherwise we can not get the elegant ``entropy sum'' relation with only the background field constant dependence.

{\bf Firstly, in odd dimensions}, the horizon function
Eq.(\ref{V(r)1}) is a function of $r^2$, which results in some
pairs of root (even number) $r_i$ and $-r_i$. On the other hand,
the entropy $S_i$ Eq.(\ref{entropy1}) is an odd order function of
$r_i$, which deriving a pair of  entropy vanishing, i.e.
$S(r_i)+S(-r_i)=0$. Hence $\Sigma_i S_i=0$ with arbitrary
cosmological constant, the ``entropy sum''  of odd dimensional
(A)dS black hole in Einstein-Maxwell gravity is vanishing. Then in
what follows of this section, we will focus on even dimensional
black holes with the $2(d-2)$ horizons. To have a further look at
the ``entropy sum'' in other special case of the horizon function,
we will present the uncharged black hole as another example. And
we will test the ``entropy sum'' in both the four and six
dimensional charged and neutral (A)dS black hole separately.

{\bf In d=4 dimensions}, the horizon function Eq.(\ref{V(r)1}) and entropy Eq.(\ref{entropy1}) reduce respectively to
\begin{align}
  &V(r)=k-\frac{2M}{r}+\frac{Q^2}{r^2}-\frac{\Lambda}{3}r^2,
\label{V(r)11}\\
  &S_i=\pi r_i^{2}.
  \label{entropy11}
\end{align}
We focus on the black holes with $4$ horizons, including the ``virtual'' horizon. To obtain the
``entropy sum'' of all horizons, we note that there is an exact equality
\begin{align}
  \sum_{i=1}^D r_i^{2}=\left(\sum_{i=1}^Dr_i\right)^2-2\left(\sum_{1\leq i<j\leq D}r_ir_j\right).
  \label{eq1}
\end{align}
where $D$ is the number of the horizons and $D=4$ for the charged
black hole. We find that it is not all necessary to list the
roots. We can conveniently apply the Vieta theorem on the horizon
function Eq.(\ref{V(r)11}), which shows relations between the
roots and the coefficients of a polynomials as follow:
\begin{align*}
  \sum_{i=1}^Dr_i=0 , \quad \sum_{1\leq i<j\leq D}r_ir_j=-\frac{3k}{\Lambda},
\end{align*}
which therefor results in
\begin{align}
  \sum_{i=1}^D r_i^{2}=\frac{6 k}{\Lambda}.
  \label{Rsum1}
\end{align}
We immediately get the ``entropy sum'' of all horizons
\begin{align}
  \sum_{i=1}^DS_i=\pi\left(\sum_{i=1}^D r_i^2\right)=\frac{6\pi k}{\Lambda}.
  \label{Ssum1}
\end{align}

When we consider the neutral black hole, the horizon function
Eq.(\ref{V(r)1}) reduces to
\begin{align}
  &V(r)=k-\frac{2M}{r}-\frac{\Lambda}{3}r^2,
\label{V(r)111}
\end{align}
which shows the number of horizons that we focus on is $D=3$, including
the ``virtual'' horizon, and the same relations between the roots
and the coefficients, hence finally we get the same ``entropy sum''
Eq.(\ref{Ssum1}) as the charged black hole case.

Thus the same ``entropy sum'' for both the four dimensional charged and neutral
black holes Eq.(\ref{Ssum1}) depends only on $\Lambda$ and $k$. They
both are independent of $M$ and $Q$. One need note that, when
$k\Lambda<0$, there exist several ``complex horizons'' for certain,
which is un-physical, for this reason that the sum of $r_i$ and
``entropy sum'' are negative. When $k=0$, the ``entropy sum'' of all
flat horizons is vanishing and there are also some un-physical
``complex horizons''. However, one should remove the possibility of
the case with no horizon, which is out of our discussion.

{\bf In d=6 dimensions}, the horizon function Eq.(\ref{V(r)1}) reduces to
\begin{align}
  &V(r)=k-\frac{2M}{r^3}+\frac{Q^2}{r^6}-\frac{\Lambda}{10}r^2,\quad \text{for charged black hole;}
\label{V(r)12}\\
  &V(r)=k-\frac{2M}{r^3}-\frac{\Lambda}{10}r^2,\quad \quad \quad \quad \text{for neutral black hole,}
\label{V(r)122}
\end{align}
and entropy Eq.(\ref{entropy1}) has possessed the form
\begin{align}
  &S_i=\frac{2}{3}\pi^2 r_i^{4}.
  \label{entropy12}
\end{align}
The number of horizons we focus on is respectively $D=8$ for charged black hole and $D=5$ for neutral black hole, including the ``virtual'' horizon. For both cases, we note a similarly exact equality
\begin{align}
  3\sum_{i=1}^D r_i^{4}=&4 \left(\sum_{i=1}^Dr_i\right) \left(\sum_{i=1}^Dr_i^3\right)+6\left(\sum_{1\leq i<j\leq D}^{D}r_ir_j\right)^2-\left(\sum_{i=1}^Dr_i\right)^4 \notag \\
  &-12\left(\sum_{1\leq i<j<m<n\leq D}r_i r_j r_m r_n\right),
\label{eq2}
\end{align}
where the Vieta theorem on the horizon function Eq.(\ref{V(r)12}) and Eq.(\ref{V(r)122}) derives the following same relations
\begin{align*}
  &\sum_{i=1}^Dr_i=0,  \quad \sum_{1\leq i<j<m<n\leq D}r_i r_j r_m r_n=0,\\
  &\sum_{1\leq i<j\leq D}^{D}r_ir_j=-\frac{10k}{\Lambda},
\end{align*}
Thus we get
\begin{align}
  \sum_{i=1}^D r_i^{4}=\frac{200 k^2}{\Lambda^2},
  \label{Rsum2}
\end{align}
which immediately shows the ``entropy sum'' of all horizons
\begin{align}
  \sum_{i=1}^DS_i=\frac{2}{3}\pi^2\left(\sum_{i=1}^D r_i^4\right)=\frac{400 k^2\pi^2}{3\Lambda^2}.
  \label{Ssum2}
\end{align}
The ``entropy sum'' for the six dimensional charged and neutral
black holes Eq.(\ref{Ssum2}) also depends only on the $\Lambda$ and $k$,
and is independent of the mass $M$ and charge $Q$. When $k=0$, the
``entropy sum'' of all flat horizons is vanishing and there are
some un-physical ``complex horizons''. When $k=\pm1$, we obtain a
positive ``entropy sum'', which seems to describe a physical
system. In addition, the ``entropy sum'' in AdS spacetime is the
same as that in the dS spacetime. Also, the spacetime with the
spherical and hyperbolic horizons shares the same ``entropy sum''.

To summarize, including the necessary effect of the un-physical
``virtual'' horizon, we find the ``entropy sum'' is dependent only on the
cosmological constant and the topology of the horizons in four and
six dimensions, while it is always vanishing in odd dimensions.
Besides, the conserved charges $M$ (black hole mass) and $Q$ (charge from
Maxwell field) play no role in the ``entropy sum'' relation. This
agrees with the result as shown in \cite{Wang:2013smb}.

\section{``Entropy sum''  of (A)dS black hole in $f(R)$ gravity}
{\bf In four dimensions}, let us first consider the action for $f(R)$ gravity with Maxwell term,
\begin{align}
  \mathcal{L}=\frac{1}{16\pi G}\int d^4 x \sqrt{-g}[R+f(R)-F_{\mu\nu}F^{\mu\nu}],
\end{align}
where $f(R)$ is an arbitrary function of the scalar curvature.
The static, spherically symmetric and with constant curvature ($R=R_0$) solutions is given by \cite{Moon:2011hq,Hendi:2011eg,Cembranos:2011sr}
\begin{align*}
  d s^2=-V(r) d t^2+\frac{d r^2}{V(r)}+r^2 d\Omega^2_{2},
\end{align*}
with the horizon function
\begin{align}
  V(r)=k-\frac{2\mu}{r}+\frac{q^2}{r^2}\frac{1}{(1+f^{\prime}(R_0))}-\frac{R_0}{12}r^2
\label{V(r)21}
\end{align}
Here the $d\Omega^2_{2}$ represents the line element of a $2$-dimensional maximal symmetric
Einstein space with constant curvature $2k$, and $k=1, 0$ and $-1$, corresponding to the spherical, Ricci flat and hyperbolic topology of the black hole horizon, respectively. The cosmological constant of this theory is then having the form $\Lambda_f=\frac{R_0}{4}$. The parameters $\mu$ and $q$ are respectively related to the mass and charge of black hole.
The number of horizons we focus on is $D=4$ for charged black hole, including the ``virtual'' horizon. For the neutral black hole, the horizon function reduces to
\begin{align}
  V(r)=k-\frac{2\mu}{r}-\frac{R_0}{12}r^2
\label{V(r)211}
\end{align}
with the number of horizons we focus on is $D=3$. For both cases, we note that the entropy of all horizons are
\begin{align}
  &S_i=\frac{A_i}{4G}(1+f^{\prime}(R_0)).
  \label{entropy21}
\end{align}
where $f^{\prime}(R_0)=\left.\frac{\partial f(R)}{\partial R}\right|_{R=R_0}$ and $A_i=4\pi r_i^2$. We find the equality Eq.(\ref{eq1}) still holds with the different $D$ here. Then with the help of the Vieta theorem on the horizon function Eq.(\ref{V(r)21}) and Eq.(\ref{V(r)211}), we obtain the following relations
\begin{align*}
  \sum_{i=1}^Dr_i=0 , \quad \sum_{1\leq i<j\leq D}r_ir_j=-\frac{12k}{R_0},
\end{align*}
which results in
\begin{align}
  \sum_{i=1}^D r_i^{2}=\frac{24k}{R_0}=\frac{6k}{\Lambda_f}.
  \label{Rsum3}
\end{align}
One finds this ``area sum'' is general in both the Einstein gravity and $f(R)$ gravity in four dimensions as shown in Eq.(\ref{Rsum1}) and Eq.(\ref{Rsum3}) respectively. In the $f(R)$ gravity, the entropy is modified as shown in Eq.(\ref{entropy21}), which immediately shows the ``entropy sum'' of all horizons
\begin{align} \label{Ssum3}
  \sum_{i=1}^DS_i=\frac{6k\pi}{\Lambda_f G}(1+f^{\prime}(R_0)).
\end{align}
This is different from the ``entropy sum'' of four dimensionally charged (A)dS black hole, because of the modification from $f(R)$ gravity. But one can also find that the ``entropy sum'' is dependent only on the $\Lambda$ and $k$. When $f(R)=\text{constant}$, the theory reduces to Einstein gravity and the ``entropy sum'' is equal to Eq.(\ref{Ssum2}). Again, when $k\Lambda<0$ and $k=0$, there are some ``virtual'' horizons for certain, because of the negative and vanishing ``area sum'', respectively. One need note that the spacetime with $\Lambda=0$ is out of the present discussion of this paper, and the ``charge-independence'' of ``entropy sum''  in that case fails as has been explicitly shown in our previous work \cite{Wang:2013smb}.

{\bf In higher dimensions}, since the standard
Maxwell energy-momentum tensor is not traceless, people have failed to derive higher
dimensional black hole/string solutions from $f(R)$ gravity coupled to standard Maxwell field. Thus the higher dimensional charged solutions are obtained only in the case of power-Maxwell
field with $d =4p$, where $p$ is the power of conformally invariant Maxwell Lagrangian \cite{Sheykhi:2012zz}. This higher dimensional solution is too complicated to continue our discussion about the ``entropy sum'' of horizon. To take a further study, one can turn to the higher-dimensional $f(R)$ gravity without Maxwell field, whose action reads as
\begin{align}
  \mathcal{L}=\int d^d x \sqrt{-g}[R+f(R)].
\end{align}
We consider the same metric form as in Eq.(\ref{metric}) with the horizon function as
\begin{align}
  V(r)=k-\frac{2m}{r^{d-3}}-\frac{R_0}{d(d-1)}r^2,
\label{V(r)22}
\end{align}
which is derived from the solutions \cite{Sheykhi:2012zz} with vanishing charge $q$. Here $m$ is an integration constant which is related to the mass of the solution. The cosmological constant in this case is $\Lambda_f=\frac{d-2}{2d}R_0$. The entropy has the same form as that of four dimensional charged $f(R)$ black hole as shown in Eq.(\ref{entropy21}) with $G=1$ and $A_i=\frac{2\pi^{(d-1)/2}}{\Gamma\left(\frac{d-1}{2}\right)}r_i^{d-2}$.

{\bf In odd dimensions}, the horizon function Eq.(\ref{V(r)22}) is a function of $r^2$ and the entropy $S_i$ is an odd order function of $r_i$. This also results in some pairs of root $r_i$ and $-r_i$ and a pair of vanishing entropy, i.e. $S(r_i)+S(-r_i)=0$. Hence $\Sigma_i S_i=0$, the ``entropy sum''  of odd dimensional (A)dS black hole in $f(R)$ gravity is vanishing.

{\bf In six dimensions}, the horizon function Eq.(\ref{V(r)22}) and area of horizons reduce respectively to
\begin{align}
  V(r)=k-\frac{2m}{r^{3}}-\frac{R_0}{30}r^2.
\label{V(r)221}
\end{align}
and $A_i=\frac{2\pi^{5/2}}{\Gamma\left(\frac{5}{2}\right)}r_i^{4}.$ The cosmological constant is $\Lambda_f=\frac{1}{3}R_0$. The number of horizons we focus on is $D=5$. We find the equality Eq.(\ref{eq2}) still holds with the different $D$ here. Then using the Vieta theorem on the horizon function Eq.(\ref{V(r)221}), we obtain
\begin{align*}
  &\sum_{i=1}^Dr_i=0,  \quad \sum_{1\leq i<j<m<n\leq D}r_i r_j r_m r_n=0,\\
  &\sum_{1\leq i<j\geq D}^{D}r_ir_j=-\frac{30k}{R_0},
\end{align*}
Thus we get
\begin{align}
  \sum_{i=1}^D r_i^{4}=\frac{1800 k^2}{R_0^2}=\frac{200 k^2}{\Lambda_f^2},
  \label{Rsum4}
\end{align}
which shows that the ``area sum'' is general in both Einstein gravity and $f(R)$ gravity in six dimensions as shown in Eq.(\ref{Rsum2}) and Eq.(\ref{Rsum4}). The ``entropy sum'' of all horizons is
\begin{align}
  \sum_{i=1}^DS_i=\frac{400 k^2\pi^2}{3\Lambda_f^2}(1+f^{\prime}(R_0)).
\end{align}
Although this is different from the ``entropy sum'' of six dimensional charged (A)dS black hole, for the modification from $f(R)$ gravity, it has the same $\Lambda_f$-dependence and $k$-dependence, and $M$-independence. When $f(R)=\text{constant}$, the theory reduces to Einstein gravity and the ``entropy sum'' is equal to Eq.(\ref{Ssum2}). There are some un-physical ``complex horizons'' when $k=0$, as the ``entropy sum'' of all flat horizons is vanishing. Still, the ``entropy sum'' in AdS and dS spacetime, and in the spacetime with the spherical and hyperbolic  horizons share the same ``entropy sum'', respectively.

To summarize briefly, including the necessary effect of the un-physical
``virtual'' horizon, we find that the ``area sum'' in $f(R)$ gravity behaviors the same as that in Einstein gravity. Besides, the ``entropy sum'' depends on the cosmological constant and the topology of the horizons, does not depend on the conserved charges $M$ and $Q$, in four and six dimensions, while they are always vanishing in odd dimensions.

\section{``Entropy sum'' of (A)dS black hole in Gauss-Bonnet gravity}
The action of the $d$-dimensional Einstein-Gauss-Bonnet-Maxwell-(A)dS theory has the form
\begin{align}
\mathcal{L}=\frac1{16\pi}\int d^d x \sqrt{-g}[R-2\Lambda+\alpha (R_{\mu\nu\gamma\delta}R^{\mu\nu\gamma\delta}-4R_{\mu\nu}R^{\mu\nu}+R^2)-4\pi F_{\mu\nu}F^{\mu\nu}],
\label{action}
\end{align}
where $\alpha$ is the Gauss-Bonnet coupling constant and the cosmological constant is $\Lambda=\pm\frac{(d-1)(d-2)}{2l^2}$ for (A)dS spacetime, $F_{\mu\nu}$ is the Maxwell field strength. The $d$-dimensional static charged Gauss-Bonne-AdS black hole solution for the above action is described by the same metric form as Eq.(\ref{metric}) with the horizon function is \cite{Boulware:1985wk,Wiltshire:1985us,Cai:2001dz,Cvetic:2001bk,Dehghani:2004vn,Cai:2013qga}
\begin{equation}
V(r)=k+\frac{r^2}{2\tilde{\alpha}}\left (1-\sqrt{1+\frac{64\pi\tilde{\alpha} M}{(d-2) r^{d-1}}-\frac{2\tilde{\alpha} Q^2}{(d-2)(d-3)r^{2d-4}}+\frac{8\tilde{\alpha}\Lambda}{(d-1)(d-2)}} \right ),
\label{V(r)3}
\end{equation}
where $\tilde{\alpha}=(d-3)(d-4)\alpha$, $M$ is the black hole mass, $Q$ is related to the charge of the black hole. The entropy has the form as
\begin{align}
  S=\frac{ r^{d-2}_+}{4}\left (1+\frac{2(d-2) k\tilde{\alpha}}{(d-4)r_+^2}\right )
\end{align}

{\bf In odd dimensions}, once again, the horizon function Eq.(\ref{V(r)3}) is a function of $r^2$ and the entropy $S_i$ is an odd order function of $r_i$, which result in some pairs of root $r_i$ and $-r_i$ and a pair of vanishing entropy, i.e. $S(r_i)+S(-r_i)=0$. Hence $\Sigma_i S_i=0$, the ``entropy sum''  of odd dimensional (A)dS black hole in Gauss-Bonnet gravity are vanishing.

{\bf In six dimensions}, the horizon function Eq.(\ref{V(r)22}) reduces to
\begin{align}
&V(r)=k+\frac{r^2}{2\tilde{\alpha}}\left (1-\sqrt{1+\frac{16\pi\tilde{\alpha} M}{r^{5}}-\frac{\tilde{\alpha} Q^2}{6r^{8}}+\frac{2\tilde{\alpha}\Lambda}{5}} \right ), \quad\text{for charged black hole;}\\
&V(r)=k+\frac{r^2}{2\tilde{\alpha}}\left (1-\sqrt{1+\frac{16\pi\tilde{\alpha} M}{r^{5}}+\frac{2\tilde{\alpha}\Lambda}{5}} \right ), \quad\quad \quad\quad\text{for neutral black hole,}
\end{align}
which result in the horizons as the roots of the following polynomials
\begin{align}
&v(r)=12\Lambda r^8-120 r^6k-120 r^4 k^2\tilde{\alpha}+480\pi M r^3-5 Q^2, \quad\text{for charged black hole;}\label{V(r)31}\\
&v(r)=\Lambda r^5-10kr^3-10 rk^2\tilde{\alpha}+40\pi M, \quad\quad \quad\quad \quad\quad \quad\quad\text{for neutral black hole.}\label{V(r)32}
\end{align}
The number of horizons we focus on is $D=8$ for charged black hole and $D=5$ for charged black hole. The entropy of horizons have the form
\begin{align}
  S_i=\frac{ r^4_i}{4}+ k\tilde{\alpha}r_i^2
\end{align}
We find the equality Eq.(\ref{eq1}) and Eq.(\ref{eq2}) still hold with the different $D$ here. Then for both charged and neutral black hole, using the Vieta theorem on the new horizon function Eq.(\ref{V(r)31}) and Eq.(\ref{V(r)32}), we obtain
\begin{align*}
  &\sum_{i=1}^Dr_i=0,  \quad \sum_{1\leq i<j<m<n\leq D}r_i r_j r_m r_n=-\frac{10 k^2\tilde{\alpha}}{\Lambda},\\
  &\sum_{1\leq i<j\leq D}^{D}r_ir_j=-\frac{10k}{\Lambda}.
\end{align*}
Thus we get
\begin{align}
  &\sum_{i=1}^D r_i^{4}=\frac{200k^2}{\Lambda^2}+\frac{40 k^2\tilde{\alpha}}{\Lambda}, \quad \sum_{i=1}^D r_i^{2}=\frac{20k}{\Lambda}.
\end{align}
which lead to the ``entropy sum'' of all horizons
\begin{align}
  \sum_{i=1}^DS_i=\frac{1}{4}\left(\sum_{i=1}^Dr^4_i\right)+ k\tilde{\alpha}\left(\sum_{i=1}^Dr^2_i\right)=\frac{50k^2}{\Lambda^2}+\frac{30 k^2\tilde{\alpha}}{\Lambda}.
\end{align}
Obviously, this is different from the ``entropy sum'' of six dimensionally charged (A)dS black hole in Einstein gravity and $f(R)$ gravity. But the ``entropy sum'' has the same $\Lambda$-dependence and $k$-dependence, and $M$-independence and $Q$-independence. Besides, It also depends on $\alpha$. When $\alpha=0$, the theory reduces to Einstein gravity and the ``entropy sum'' is proportional to Eq.(\ref{Ssum2}). There are some un-physical ``complex horizons'' when $k\Lambda<0$ and $k=0$, as the  sum of $r_i^2$ of all flat horizons is negative and zero respectively. However, this ``entropy sum'' in AdS and dS spacetime, and in the spacetime with the spherical and hyperbolic  horizons do not share the same ``entropy sum'', which is different from that of Einstein gravity and the $f(R)$ gravity.

Finally, to give a brief summary, including the necessary effect of the un-physical
``virtual'' horizon, we conclude that the ``entropy sum''  depends on the cosmological constant, the Gauss-Bonnet coupling constant and the topology of the horizons, while it does not depend on the conserved charges $M$ and $Q$, in six dimensions. The extra background field constant dependence also appears in the spacetime with other matter source as shown in \cite{Wang:2013smb}. Again, the ``entropy sum''  is vanishing in odd dimensions.

\section{``Entropy sum'' of black hole in gauged supergravity theory}
``Entropy sum'' relation is also valid in supergravity theory. Let us check it explicitly.

{\bf In the four dimensions}, a gauged pairwise equal charges solution which is constructed in
\cite{Chong:2004na},
has the metric
\begin{align} \label{eq:D=4 gauged pairwise}
ds_4^2&=-\frac{\Delta_r}{W}\left(dt-\frac{a\sin^2\theta}{\Xi} d\phi\right)^2
       +W\left(\frac{dr^2}{\Delta_r}+\frac{d\theta^2}{\Delta_\theta}\right)\notag \\
       &+\frac{\Delta_\theta\sin^2\theta}{W}\left(adt-\frac{r_1r_2+a^2}{\Xi}d\phi\right)^2,
\end{align}
where
\begin{align*}
&\Delta_r=r^2+a^2-2mr+g^2r_1r_2(r_1r_2+a^2), \\
&\Delta_\theta=1-g^2a^2\cos^2\theta, \\
&W=r_1r_2+a^2\cos^2\theta,
\end{align*}
and
\[r_1=r+2m\sinh^2\delta_1 \qquad r_2=r+2m\sinh^2\delta_2. \]
As discussed in \cite{Chow:2008ip}, when the parameter $\delta_i=0$, the metric (\ref{eq:D=4 gauged pairwise})
should reduce to the four dimensional Kerr-AdS metric\footnote{The metric in \cite{Chong:2004na} is not suitable
here.}. Solve the equation $\Delta_r(r)=0$ to get the four horizons. And the entropy for each horizon is
$S(r_\alpha)=\frac{A(r_\alpha)}{4}$, where the area of each horizon is
\begin{equation}
A(r_\alpha)=\frac{4\pi (r_{1\alpha}r_{2\alpha}+a^2)}{\Xi}.
\end{equation}
With no difficulty we get
\begin{equation}
\sum_{\alpha=1}^4A(r_\alpha)=-\frac{8\pi}{g^2},
\end{equation}
and the entropy sum
\begin{equation} \label{eq:D=4 entropy sum}
\sum_{\alpha=1}^4S(r_\alpha)=-\frac{2\pi}{g^2}.
\end{equation}

{\bf In the six dimensions}, a gauged charged rotating black hole is considered in \cite{Chow:2008ip}. The
metric can be presented concisely in Jacobi-Carter coordinates. However, when we concern with thermodynamic
quantities, we use angular velocities measured with respect to a non-rotating frame at infinity and move to
Boyer--Lindquist time and azimuthal coordinates, which is discussed intensively in \cite{Chow:2008ip}. For
simplification, we do not intend to write the details here. The horizon function is
\begin{align}
f(r)=(r^2 + a^2)(r^2 + b^2)+g^2[r(r^2+a^2)+q][r(r^2+b^2)+q]-2mr,
\end{align}
and the entropy for each horizon is
\begin{align}
S(r_i)=\frac{2\pi^2\left((r_i^2+a^2)(r_i^2+b^2)+qr_i\right)}{3\Xi_a\Xi_b}.
\end{align}
Rewrite $f(r)=0$ to a polynomial function
\begin{align}
&g^2 r^6+ ( 1+{g}^{2}{a}^{2}+{g}^{2}{b}^{2}){r}^{4}+2\,{g}^{2}q{r}^{3}+(g^2 a^2 b^2+a^2+b^2)r^2 \notag \\
&+ (-2\,m+g^2 a^2 q+g^2 b^2 q)r+a^2 b^2+g^2 q^2=0.
\end{align}
It has six roots at most, which correspond to six horizons, including un-physical ones. However, we do not need to solve them analytically. By using the Vieta theorem, it is not difficult to compute
\begin{align} \label{eq:D=6 entropy sum}
\sum_{i=1}^6 S(r_i)=\frac{4\pi^2}{3g^4}.
\end{align}
Both (\ref{eq:D=4 entropy sum}) and (\ref{eq:D=6 entropy sum}) only relate to the gauge parameter $g$, which can be interpreted as an effective cosmological constant in the position.

{\bf In the odd dimensions}, unlike the Einstein-Maxwell, $f(R)$ or Gauss-Bonnet theory, there is not a
universal metric in arbitrary $d$ dimensions. So we have to check explicitly. In $d=5$ \cite{Chong:2005hr} and
$d=7$ \cite{Chow:2008}, the ``entropy sum'' equals zero. It is the same as our conclusion above.

In these solutions, we note that there may exist more than one charge \cite{Chong:2004na} or angular momentum
\cite{Chow:2008ip}. However, they play no role in the entropy sum, which suggests again that the entropy sum is
``universal'' as we have anticipated.

\section{Conclusions}
As we have seen above, the ``entropy sum'' relation of (A)dS
charged black holes in higher dimensional Einstein-Maxwell
gravity, $f(R)$ gravity depends only on the cosmological constant.
For the case in Gauss-Bonnet gravity, there are the cosmological
constant dependence and Gauss-Bonnet coupling constant dependence
holding together. We have taken the detail discussions in four,
six and all odd dimensions, putting all the cases together, we
conclude that the ``entropy sum'' shares the following properties:
\begin{enumerate}
\item It has got the cosmological constant dependence (and
Gauss-Bonnet coupling constant dependence for Gauss-Bonnet gravity) in four and six
dimensions. They are also dependent of the topology of the
horizons in four and six dimensions;

\item It is always vanishing in odd dimensions;

\item It involves all possible bifurcating horizons including the ``virtual'' horizon, i.e. we consider all roots of the horizon function. Generically, this involves complex roots, and hence un-physical horizons. The same phenomenon has appeared in the discussion of universal entropy relation in Schwarzschild-de Sitter and Reissner-Nordstr{\"o}m-de Sitter black holes \cite{Visser:2012wu};

\item The conserved charges $M$ (the mass), $Q$ (the charge from Maxwell field or supergravity) and the parameter $a$ (the angular momentum) play no role in the ``entropy sum'' relation.
\end{enumerate}

Generically, one may expect that the background constant dependence and the horizon geometry dependence do still hold in even dimensions. In addition, the solutions discussed in this paper
all include a cosmological constant, with the reason that the cases in
asymptotical flat spacetime is a failed example
(One can see the appendix for the detail calculation to this argument). Thus it is helpful to understand the entropy relation further.

Actually, the entropy sum are expected able to shed some light on the understanding of the microscopics physics of black holes, possibly like the entropy product case. Note the mass-independence of entropy product failed in static,neutral asymptotical (A)dS spacetime \cite{Faraoni:2012je,Detournay:2012ug,Visser:2012wu}, while the mass-independence of entropy sum always holds in asymptotical (A)dS spacetime. Hence for the cases in asymptotical (A)dS spacetime, the ``entropy sum'' are expected to play the role as the ``entropy product'' doing in asymptotical flat spacetime. Though this is still not clear, we can present some applications of the entropy sum here. It is found that the entropy sum may lead to some Penrose-like inequalities in both asymptotically (A)dS and flat spacetime. We firstly take the entropy sum of Kerr black hole as an example for asymptotically flat spacetime. There are only two horizons: the event horizon $r_E$ and Cauchy horizon $r_C$. The result is in \cite{Wang:2013smb}, which is obviously violating the mass independence relation.
\begin{align}
  S_E+S_C=4\pi\,M^2,
\end{align}
where $S_E$ and $S_C$ are the entropy of event horizon and Cauchy horizon, respectively. $M$ is the mass of Kerr black hole. As the event horizon and Cauchy horizon entropy are both positive, one can easily find
\begin{align}
  \sqrt{\frac{S_E}{4\pi}}\leq\,M,
\end{align}
or
\begin{align}
  \sqrt{\frac{A_E}{16\pi}}\leq\,M,
\end{align}
which is the exact Penrose inequality of black holes. Here $A_E$ is the area of the event horizon. This is the first example of geometrical inequality for black holes. One can see the recent review article \cite{Mars:2009cj} and references therein for discussion details. Then we focus on the asymptotically (A)dS spacetime and consider the case of Schwarzschild-dS black holes \cite{Wang:2013smb} for simplicity. There are three horizons: the event horizon $r_E$, cosmological horizon $r_C$ and ``virtual'' horizon $r_V$. Note the final one has a negative value and is not physical. The entropy sum is \cite{Wang:2013smb}
\begin{align}
  S_E+S_C+S_V=\frac{6\pi}{\Lambda},
\end{align}
where $S_E$, $S_C$ and $S_V$ are the entropy of event horizon, cosmological horizon and ``virtual'' horizon, respectively. Similarly, the area entropy indicates that they are all positive, which leads to the following Penrose-like inequality
\begin{align}
S_E\leq\frac{6\pi}{\Lambda},
\end{align}
or
\begin{align}
\sqrt{\frac{A_E}{24\pi}}\leq\,L,
\end{align}
which is the equivalent geometrical form. Here, $L$ is the cosmological radius and $A_E$ is the area of the event horizon. The two examples are interesting, clear but simple to discuss. It is very interesting to consider the more general cases by using the entropy sum.

We see clearly in the above Kerr black hole discussion, the asymptotically flat case that the mass-independence relation for the entropy sum violated. We can get more examples in \cite{Castro:2013pqa,Wang:2013smb,Xu:2014qaa}. Also, we have left out the complicated case by considering the multi-rotations in higher dimensions, while the ``entropy sum'' does not depend on the rotation in four dimensional Einstein gravity and Einstein-Weyl gravity. All these can be left as future possible efforts.

\section*{Acknowledgments}
Enlightening discussions with Professor M. Cvetic is highly appreciated. Also, we thank the referees for their very helpful comments to make this work improved greatly. Wei Xu is supported by the Research Innovation Fund of Huazhong University of Science and Technology (2014TS125). This work is partially supported by the Natural
Science Foundation of China (NSFC) under Grant No.11075078 and the State Key Laboratory of ITP-CAS fund.

\appendix
\section{``Entropy sum'' in asymptotical flat spacetime - the failure cases}
In this appendix, we present some cases of the entropy sum in asymptotical flat spacetime as examples of the failure cases. One can find that the cosmological constant dependence and mass independence do not hold.
\subsection{``Entropy sum'' of Kerr black hole}
We firstly revisit the entropy sum of Kerr black hole \cite{Wang:2013smb}. The Kerr black hole takes the following form
\begin{align*}
  &ds^2=\frac{\Sigma}{\Delta}dr^2-\frac{\Delta}{\Sigma}(dt-a\sin^2\theta\,d\phi)^2+\Sigma\,d\theta^2+\frac{\sin^2\theta}{\Sigma}\bigg((r^2+a^2)d\phi-adt\bigg)^2\\
  &\Sigma=r^2+a^2\cos^2\theta
\end{align*}
with the horizon function
\begin{align*}
  \Delta=r^2-2M r+a^2,
\end{align*}
where parameters $M$ and $a$ are related to the mass and the angular momentum. There are two horizons: the event horizon $r_E=M+\sqrt{M^2-a^2}$ and the Cauchy horizon $r_C=M-\sqrt{M^2-a^2}$. Introducing the area of each horizon $A(r_i)=4 \pi (r_i^2+a^2)$, we obtain the entropy sum \cite{Wang:2013smb}
\begin{align*}
S_E+S_C=\frac{A_E}{4}+\frac{A_C}{4}=4\pi\,M^2.
\end{align*}
Apparently it is mass dependent and the ``universal'' property of ``entropy sum'' breaks down. Actually, generalizing to the Kerr-Newman black hole, this result still works. We can easily calculate the entropy sum
\begin{align*}
S_E+S_C=\frac{A_E}{4}+\frac{A_C}{4}=2\pi\bigg(2\,M^2-q^2\bigg).
\end{align*}
\subsection{``Entropy sum''  of flat black hole in reduced $f(R)$ gravity}
The ``flat'' black hole in four dimensional $f(R)$ gravity has the reduced horizon function
\begin{align*}
  V(r)=k-\frac{2\mu}{r}+\frac{q^2}{r^2}\frac{1}{(1+f^{\prime}(R_0))},
\end{align*}
where the cosmological constant $\Lambda_f=\frac{R_0}{4}$ in Eq.(\ref{V(r)21}) is vanishing. There are only two horizon as well: the event horizon $r_E$ and the Cauchy horizon $r_C$. For this case,
the corresponding relations from Vieta theorem reduce to
\begin{align*}
  r_E+r_C=\frac{2\mu}{k}, \quad r_Er_C=\frac{q^2}{k(1+f^{\prime}(R_0))}.
\end{align*}
Then inserting the entropy of horizons Eq.(\ref{entropy21}) and applying the exact equality Eq.(\ref{eq1}), we can obtain the ``entropy sum'' of all horizons
\begin{align*}
  S_E+S_C=\frac{2\pi}{Gk^2}\left(2\mu^2(1+f^{\prime}(R_0))-kq^2\right).
\end{align*}
Again we find that the ``entropy sum'' of $f(R)$ black holes in asymptotical flat spacetime depend on the conserved charges $M$ and $Q$ ($\mu$ and $q$).

\vspace{10pt}
To summarize briefly, the entropy sum in asymptotical flat spacetime always depends on the mass and electric charge, hence the ``universal'' property of ``entropy sum'' fails. The same phenomenon also happen
in the study of entropy product \cite{Visser:2012wu,Castro:2013pqa}, in which the mass independence of entropy product breaks down for Schwarzschild-(A)dS black hole and Lovelock black hole.

\providecommand{\href}[2]{#2}\begingroup
\small\itemsep=0pt
\providecommand{\eprint}[2][]{\href{http://arxiv.org/abs/#2}{arXiv:#2}}

\end{document}